\documentclass[12pt]{article}

\usepackage[letterpaper,margin=1in,includeheadfoot]{geometry}
\usepackage{amsmath,amssymb,amsfonts,bm,booktabs,microtype,setspace,natbib}
\usepackage[hidelinks]{hyperref}
\usepackage{graphicx}
\usepackage{amsthm}
\usepackage{authblk}

\newtheorem{theorem}{Theorem}
\title{A Bayesian Edge-Space Framework for Whole-Connectome Inference in Multisite Autism Neuroimaging}

\author[1]{Montserrat Fuentes}
\author[2]{Veronica B. Patterson}

\affil[1]{Department of Mathematics, St. Edward's University, Austin, Texas, USA}
\affil[2]{Department of Statistics, Rice University, Houston, Texas, USA}

\date{}

\begin{document}
\maketitle
\begin{center}
\textbf{Corresponding author:} Veronica B. Patterson\\
Department of Statistics, Rice University\\
Houston, TX 77005, USA\\
\href{mailto:vp30@rice.edu}{vp30@rice.edu}
\end{center}

\noindent\textbf{Running head:} Bayesian Edge-Space Connectome Inference

\begin{abstract}
Autism spectrum disorder (ASD) is associated with heterogeneous alterations across distributed brain systems, creating challenges for whole-connectome inference. The difficulty arises not only from the large number of connections, but also from dependence among effects indexed by anatomically and functionally related region pairs. We introduce a Bayesian Edge-Space regression framework that treats each participant's connectome as a network-valued response and models the adjusted ASD effect over unordered brain-region pairs.  The main methodological contribution is a positive-semidefinite covariance construction defined directly on connections. Anatomical and diagnosis-blind functional similarities are lifted from regions to edge space through a symmetrized endpoint-matching operation that preserves endpoint identity and is invariant to endpoint ordering. An additive Bayesian hierarchy estimates anatomical, functional, and interaction contributions together with multisite adjustments and connection-specific effects.  Theoretical results establish covariance validity and continuous nesting of the structured components. Low-rank kernel representations and an exact sufficient-statistic reduction enable whole-connectome computation without preliminary edgewise estimation. Simulations show improved recovery of the effect surface, particularly under weak signals.  In the Autism Brain Imaging Data Exchange, the framework identifies widespread reductions together with localized increases in ASD-associated connectivity. This pattern supports heterogeneous reorganization across distributed neural systems rather than uniform hyper- or hypoconnectivity. Under the fitted parameterization, the functional component has the largest structural scale, indicating organization beyond anatomical proximity alone. 
\end{abstract}


\section{Introduction}
\label{sec:introduction}

Autism spectrum disorder (ASD) is a heterogeneous neurodevelopmental condition whose neural basis is distributed across interacting systems rather than localized to a single brain region. Resting-state functional magnetic resonance imaging (rs-fMRI) provides a noninvasive view of this organization by measuring spontaneous activity across anatomically defined regions. Pairwise associations among regional signals form a functional connectome, representing each participant by a weighted network and providing a natural object for studying autism as altered communication across distributed neural systems \citep{BullmoreSporns2009,Hull2017}.

The Autism Brain Imaging Data Exchange (ABIDE) broadened this research by assembling imaging and phenotypic information from multiple international centers \citep{DiMartino2014}. Its multisite design improves the potential generalizability of findings but creates substantial statistical difficulty. Functional connectivity is high dimensional, connections are strongly dependent, and scanner, protocol, recruitment, demographic, and motion differences may be comparable in magnitude to the biological signal. Whole-connectome inference must estimate adjusted ASD effects while accounting simultaneously for dependence among connections and heterogeneity among acquisition sites \citep{Fortin2017,Johnson2007}.

For a parcellation with $R$ regions, the connectome contains $R(R-1)/2$ unique undirected connections,  creating a high-dimensional inferential problem. These connections are also dependent because they may share endpoints, link anatomically similar regions, or occupy related positions in large-scale functional organization. We refer to the domain of unordered region pairs as \emph{edge space}. The inferential task is to estimate a disease-effect function over this domain while accounting for dependence among biologically related connections.

This relationship-valued structure is not unique to brain connectivity. In many scientific settings, the effects of interest are attached to interactions, pairs, or links rather than to individual units. Examples include gene--gene interactions, ecological associations, communication networks, and linked health outcomes. These applications require methods that preserve inference for individual relationships while representing dependence across the broader relational domain.

In autism, this inferential problem is especially important because connectivity differences may be distributed across interacting systems rather than concentrated in a small number of isolated connections. Widespread decreases together with localized increases would suggest heterogeneous reorganization across distributed neural systems. Such a pattern would not support a uniform characterization of autism as either hyperconnected or hypoconnected. Determining whether these effects follow physical proximity, population functional architecture, or both requires a model that estimates each connection-specific effect and the organization of the complete effect surface. This motivates an Edge-Space covariance model that preserves local interpretation while borrowing information across anatomically and functionally related connections.

Conventional edgewise analyses fit a separate regression at every connection and then apply multiplicity adjustment \citep{BenjaminiHochberg1995}. They retain connection-specific interpretation but discard dependence. Network-based procedures recognize coordinated alteration but target thresholded components rather than the magnitude, direction, and uncertainty of an adjusted effect at every connection \citep{Zalesky2010}. Graph summaries provide interpretable measures of integration and segregation, but distinct edge-level patterns can produce similar global measures \citep{BullmoreSporns2009,RubinovSporns2010}. Predictive analyses of ABIDE focus on classifying participants as ASD or control. Our goal is different: to estimate the adjusted ASD effect at each connection and quantify its uncertainty across the complete connectome \citep{Heinsfeld2018,Parisot2018}.

Recent statistical work has developed Bayesian network-response models, matrix-response regression, covariance regression, and network-predictor methods that respect high-dimensional structure \citep{DuranteDunson2018,Wang2017,Hu2021,Zhao2021,GuhaRodriguez2021,Ju2025}. These methods establish the importance of modeling networks coherently, but their principal targets are often latent representations, prediction, or lower-dimensional covariance summaries. They do not directly address the objective considered here: estimation of an adjusted effect at every connection together with an interpretable covariance model for how those effects are organized.

We introduce a Bayesian Edge-Space regression model that treats each participant's complete connectome as a network-valued response and represents the adjusted ASD effect as a stochastic function on unordered brain-region pairs. Among its methodological contributions is a positive-semidefinite covariance construction defined directly on connections. Anatomical and diagnosis-blind functional similarities are lifted from regions to edge space through a symmetrized endpoint-matching operation that preserves endpoint identity and remains invariant to endpoint ordering. An interaction kernel captures structure supported jointly by both representations. More generally, the construction provides a way to transfer scientifically meaningful similarities from individual units to effects defined on unordered relationships.

Covariance models have also been developed for processes observed on geometric networks. For example, \citet{AnderesMollerRasmussen2020} construct isotropic covariance functions for observations indexed by vertices and points lying along Euclidean graph edges, using geodesic and resistance metrics. That setting differs from the present problem: here, the edge is itself the inferential unit, identified by an unordered pair of brain regions. Related symmetric Kronecker kernels arise in pairwise and dyadic learning \citep{StockEtAl2018,ViljanenEtAl2022}, primarily for prediction or regularized learning. The present framework instead uses endpoint-based pair kernels as covariance components in network-response regression to estimate adjusted effects and posterior uncertainty at every connection.

The Bayesian hierarchy estimates the relative contributions of the anatomical, functional, and interaction components together with the ASD surface, participant-level adjustments, site effects, and residual variation. Low-rank kernel representations and an exact sufficient-statistic reduction make whole-connectome posterior computation feasible without replacing the participant-level regression with preliminary edgewise estimates. Theoretical results establish that the lifted kernels define valid covariance matrices on unordered edges and that the structured models vary continuously as individual components shrink toward zero.

The resulting framework provides connection-specific inference while also describing how the complete disease-effect surface is organized across anatomical and functional brain architecture. In ABIDE, this supports a more precise and biologically interpretable analysis of distributed autism-associated connectivity. More broadly, the Edge-Space construction establishes a general foundation for inference when scientific effects are defined on interactions rather than individual units. Section~\ref{sec:data} introduces the ABIDE cohort and inferential target. Section~\ref{sec:methods} presents the model and covariance construction, and Section~\ref{sec:application} applies the method to multisite autism neuroimaging. The complete simulation study, theoretical details, and additional displays are provided in the Supplementary Materials.

\section{Data and Scientific Objective}
\label{sec:data}

The scientific objective is to characterize how autism spectrum disorder (ASD) is associated with functional connectivity across the brain after accounting for participant characteristics and systematic differences among imaging sites. The Autism Brain Imaging Data Exchange I (ABIDE I) is well suited to this objective because it combines resting-state functional magnetic resonance imaging (rs-fMRI) and phenotypic information from independently acquired studies conducted at multiple international centers \citep{DiMartino2014}. The resulting heterogeneity is scientifically valuable because it broadens the population and acquisition settings represented in the analysis, but it also creates the central inferential challenge addressed in this paper: disease-associated connectivity must be separated from variation attributable to scanners, protocols, recruitment practices, demographic composition, and head motion.

We used the quality-checked ABIDE I preprocessed derivatives distributed through the Preprocessed Connectomes Project. Regional time-series files were obtained programmatically using the CPAC preprocessing pipeline, the AAL regional time-series derivative, temporal band-pass filtering from 0.01 to 0.10 Hz, and no global-signal regression. The analysis used only the de-identified regional time series and phenotypic fields required for diagnosis, acquisition site, age, sex, and mean framewise displacement. Original ABIDE sites obtained their own ethics approvals and participant consent or assent; the present study is a secondary analysis of a public, de-identified research resource.

The downloaded archive contained 871 regional time-series files. Each file was required to be a two-dimensional matrix with exactly 116 regional columns, at least 50 time points, and finite entries. Participants with more than three effectively constant regional series were excluded because correlations involving nearly constant signals are undefined or unstable. Imaging and phenotypic records were then aligned, and participants lacking any variable required by the regression model were removed. The resulting analysis cohort contained 792 participants from 20 acquisition sites. ASD diagnosis is the primary explanatory variable, while age, sex, mean framewise displacement, and acquisition-site indicators enter the model as adjustment variables. The estimand is therefore an adjusted ASD contrast rather than a marginal comparison of the diagnostic groups.

\begin{figure}[p]
\centering
\makebox[\textwidth][c]{%
\includegraphics[width=1.08\textwidth]
{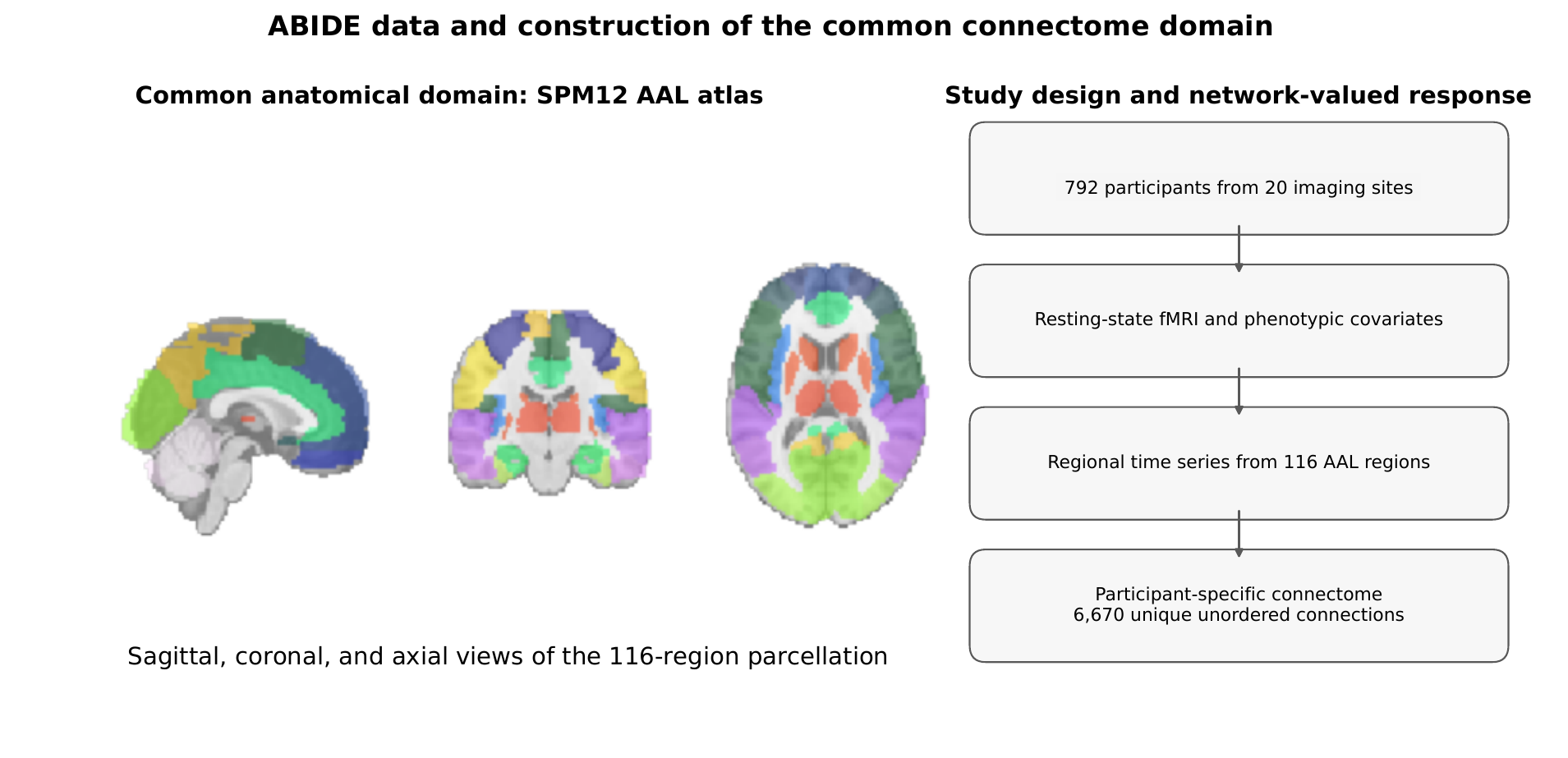}}
\caption{
ABIDE study design and common connectome domain. The analysis includes 792
participants from 20 imaging sites. Resting-state fMRI and phenotypic
information are represented using the 116-region SPM12 AAL atlas. Regional
time series are converted into participant-specific connectomes containing
6,670 unique undirected connections. This common domain provides the
network-valued response on which the proposed Edge-Space covariance model is
defined.
}
\label{fig:abide_domain}
\par\noindent\textbf{Alt text:} Flow diagram linking the ABIDE I multisite
cohort to 116 AAL regional time series and participant-specific connectomes
with 6,670 undirected edges; orthogonal atlas views show the shared anatomical
domain.
\end{figure}

Figure~\ref{fig:abide_domain} summarizes the progression from the multisite cohort to the common network-valued response and establishes the common anatomical domain.

To place every participant on the same anatomical domain, the brain was parcellated using the SPM12 implementation of the Automated Anatomical Labeling (AAL) atlas \citep{TzourioMazoyer2002}. After exclusion of the background label, the atlas contains 116 cortical and subcortical regions. The common atlas and a fixed ordering of region pairs ensure that a given response coordinate has the same biological interpretation for every participant.

For participant $i$, let $T_{ir}(t)$ denote the resting-state time series in region $r$. Functional connectivity between regions $r$ and $s$ is defined by the Fisher-transformed Pearson correlation,
\begin{equation}
Y_i(\{r,s\})
=
\operatorname{arctanh}
\left[
\operatorname{cor}
\left\{
T_{ir}(t),T_{is}(t)
\right\}
\right].
\label{eq:functional_connectivity}
\end{equation}
The Fisher transformation stabilizes the correlation scale and provides a suitable working response for regression and uncertainty quantification. Each participant's values form a symmetric $116\times116$ matrix. Because the diagonal is not part of the scientific target and the two triangles duplicate the same off-diagonal information, the complete nonredundant connectome contains
\[
q=\binom{116}{2}=6,670
\]
unique undirected connections.

Retaining the upper triangle provides a fixed computational representation but does not imply independence among its entries. Each value remains indexed by the two regions defining that connection; Web Figure 2 illustrates this representation for one participant.

We define the common edge domain
\begin{equation}
\mathcal E
=
\left\{
\{r,s\}:1\le r<s\le116
\right\},
\label{eq:data_edge_domain}
\end{equation}
where $\{r,s\}$ and $\{s,r\}$ denote the same undirected connection. Participant $i$ contributes the network-valued response
\[
\bm Y_i=\{Y_i(e):e\in\mathcal E\}.
\]
This representation preserves the entire off-diagonal connectome while locating every outcome on a relationship-valued domain. That domain is the foundation for the new covariance construction in Section~\ref{sec:methods}: anatomical and diagnosis-blind functional information defined for individual regions are lifted to similarities among unordered edges.

Let $\bm z_i$ denote age, sex, motion, and acquisition-site adjustment variables. The primary scientific estimand is the adjusted ASD effect
\begin{equation}
\beta_{\mathrm{ASD}}(e)
=
E\!\left[
Y_i(e)\mid \mathrm{ASD}_i=1,\bm z_i
\right]
-
E\!\left[
Y_i(e)\mid \mathrm{ASD}_i=0,\bm z_i
\right],
\qquad e\in\mathcal E.
\label{eq:asd_estimand}
\end{equation}
Positive values indicate stronger expected connectivity in ASD after adjustment, while negative values indicate weaker expected connectivity. The estimand is defined over all connections because autism is not expected to produce a uniform global increase or decrease in connectivity; increases and decreases may coexist across distributed neural systems \citep{Hull2017}. The inferential target is therefore the complete effect surface
\[
\bm\beta_{\mathrm{ASD}}
=
\{\beta_{\mathrm{ASD}}(e):e\in\mathcal E\},
\]
together with connection-specific uncertainty and inference on the biological organization that supports estimation of that surface.

\section{Methods}
\label{sec:methods}

Each participant contributes a complete functional connectome on the common domain of unordered region pairs. The goal is to estimate the adjusted ASD effect at every connection while using dependence among anatomically and functionally related connections. Web Figure 1 in the Supplementary Materials summarizes the computational framework.

\subsection{Regression on edge space}
\label{subsec:edge_regression}

Let $\mathcal V=\{1,\ldots,R\}$ be the common set of regions and
\[
\mathcal E=\{\{r,s\}:1\le r<s\le R\},\qquad q=|\mathcal E|=\binom{R}{2},
\]
the domain of undirected edges. For participant $i$, let $Y_i(e)$ be Fisher-transformed connectivity at edge $e$, let $\bm x_i$ contain the intercept, ASD status, age, sex, and motion, and let $s_i$ denote acquisition site. We model
\begin{equation}
Y_i(e)=\bm x_i^\top\bm\beta(e)+\gamma_{s_i}(e)+\varepsilon_i(e),
\qquad e\in\mathcal E,
\label{eq:edge_regression}
\end{equation}
with a reference-site constraint for identifiability. Under a fixed ordering of $\mathcal E$,
\begin{equation}
Y=XB+H\Gamma+E,
\label{eq:matrix_edge_regression}
\end{equation}
where $H$ is the reference-coded site design. Vectorization is computational only: every response coordinate retains its two regional endpoints. The primary estimand is the complete adjusted ASD surface $\bm\beta=\{\beta_{\mathrm{ASD}}(e):e\in\mathcal E\}$.

Each edge response is centered and scaled across participants before fitting, and posterior summaries are transformed back to the Fisher-$z$ scale. The independent, anatomy-only, and full models use the same likelihood, adjustments, and response standardization; they differ only in the covariance assigned to the ASD surface.

\subsection{Covariance construction on unordered edges}
\label{subsec:edge_covariance}

The main methodological contribution is a covariance defined directly on unordered connections. Let $\kappa$ be a positive-semidefinite kernel on regions. For $e=\{r,s\}$ and $e'=\{u,v\}$, define
\begin{equation}
k(e,e')=\frac12\{\kappa(r,u)\kappa(s,v)+\kappa(r,v)\kappa(s,u)\}.
\label{eq:unordered_edge_kernel}
\end{equation}
The two terms correspond to the valid endpoint alignments. Their average is invariant to endpoint ordering while preserving endpoint identity, unlike a midpoint representation.

For anatomy, let $a_r\in\mathbb R^3$ be the AAL centroid and use
\begin{equation}
\kappa_A(r,u)=\exp\{-\|a_r-a_u\|^2/(2\rho_A^2)\}.
\label{eq:anatomical_node_kernel}
\end{equation}
For functional organization, a diagnosis-blind population adjacency matrix is formed on the 116 regions. The next 12 eigenvectors of its symmetric normalized graph operator, after removal of the trivial leading vector, define coordinates $g_r\in\mathbb R^{12}$, and
\[
\kappa_G(r,u)=\exp\{-\|g_r-g_u\|^2/(2\rho_G^2)\}.
\]
The embedding is constructed without ASD status and treated as fixed during posterior computation. The ranges are fixed before fitting and calibrated to the empirical regional-distance scales: $\rho_A=31.0$ atlas-coordinate units and $\rho_G=0.50$ embedding units. Web Appendix C reports sensitivity to nearby values.

Applying equation~\eqref{eq:unordered_edge_kernel} gives Gram matrices $K_A$ and $K_G$ on $\mathcal E$. Their concurrence is represented by $K_{AG}=K_A\circ K_G$. The following result, proved in Web Appendix A, establishes covariance validity.

\begin{theorem}[Validity of the Edge-Space covariance]
\label{thm:edge_space_psd}
If $\kappa_A$ and $\kappa_G$ are positive-semidefinite region kernels, then $K_A$, $K_G$, and $K_{AG}$ are positive semidefinite. Hence
\begin{equation}
\Sigma_\beta=\sigma_A^2K_A+\sigma_G^2K_G+\sigma_{AG}^2K_{AG}
\label{eq:edge_space_covariance}
\end{equation}
is a valid covariance matrix on unordered edges for all nonnegative component scales.
\end{theorem}

We use the decomposition
\[
\bm\beta=\bm\beta_A+\bm\beta_G+\bm\beta_{AG},
\qquad
\bm\beta_k\sim N_q(\bm0,\sigma_k^2K_k),
\]
with independent components. The anatomy-only model sets $\sigma_G=\sigma_{AG}=0$. The independent benchmark is specified separately as $\bm\beta\sim N_q(\bm0,\tau_I^2I_q)$ and is not a limiting case of the structured family.

\subsection{Bayesian computation and inference}
\label{subsec:likelihood_inference}

Dense $q\times q$ kernels are reduced by retaining the smallest number of leading eigenvalues explaining at least 95\% of each kernel trace. The resulting low-rank kernels remain positive semidefinite. After truncation, each is normalized to average diagonal one,
\begin{equation}
K_k^*=\widetilde K_k/\{q^{-1}\operatorname{tr}(\widetilde K_k)\},
\qquad q^{-1}\operatorname{tr}(K_k^*)=1,
\label{eq:kernel_normalization}
\end{equation}
so the component scales are comparable under the fitted parameterization. The retained ranks are 360, 51, and 195 for the anatomical, functional, and interaction kernels. Writing $K_k^*=Z_k^*(Z_k^*)^\top$ gives the noncentered representation
\[
\bm\beta=\sigma_AZ_A^*\bm\eta_A+\sigma_GZ_G^*\bm\eta_G+\sigma_{AG}Z_{AG}^*\bm\eta_{AG},
\qquad \bm\eta_k\sim N(\bm0,I).
\]

The structured scales use $t_4^+(0,0.10)$ priors and the residual standard deviation uses $t_4^+(0,1)$. Intercept, age, sex, motion, and site surfaces are estimated jointly through prespecified lower-rank bases with weakly informative half-Student-$t$ scale priors. The same nuisance and site specifications are used in all model comparisons. Web Appendix B gives the complete hierarchy and computational details.

For standardized responses,
\[
Y_i^*(e)\mid B,\Gamma,\sigma_\varepsilon
\sim N\{\bm x_i^\top\bm\beta(e)+\gamma_{s_i}(e),\sigma_\varepsilon^2\}.
\]
With $W=[X\ H]$, the Gaussian likelihood depends on the data only through $W^\top W$, $W^\top Y^*$, and $\operatorname{diag}\{(Y^*)^\top Y^*\}$. This reduction is exact and avoids repeated processing of the participant-by-edge array.

Posterior sampling uses four No-U-Turn Sampler chains, each with 1,500 warmup and 1,500 retained iterations, target acceptance probability 0.97, and seed 20260720. Convergence is assessed using rank-normalized $\widehat R$, bulk and tail effective sample sizes, and divergent transitions. For each edge, we report the posterior mean, credible interval, and posterior sign probability
\begin{equation}
\pi_e=\max[P\{\beta_{\mathrm{ASD}}(e)>0\mid Y\},P\{\beta_{\mathrm{ASD}}(e)<0\mid Y\}].
\label{eq:posterior_sign_probability}
\end{equation}
The structural scales and component surfaces are interpreted jointly and conditionally on the selected, normalized kernels because the components are not required to be orthogonal.

\paragraph{Simulation evaluation.}
The complete simulation study is reported in Web Appendix C and Web Tables 1--3 of the Supplementary Materials. Across 100 Monte Carlo replications, structured Edge-Space estimation reduced RMSE and increased correlation with the true disease-effect surface relative to independent edgewise regression, with the largest gains under weak signals. A separate limited diagnostic documents the computational behavior of the full Bayesian hierarchy.

\section{Application to Multisite Autism Neuroimaging}
\label{sec:application}

We apply the proposed Bayesian Edge-Space model to resting-state functional
connectivity from the Autism Brain Imaging Data Exchange I (ABIDE I). As
described in Section~\ref{sec:data}, the analysis includes 792 participants
from 20 imaging sites, each represented by the 6,670 undirected connections
among 116 AAL regions. ASD diagnosis is the primary explanatory variable, with
age, sex, head motion, and acquisition site included as adjustment variables.

The analysis asks how ASD-associated connectivity differences are distributed
across the complete connectome and whether their organization is explained by
anatomical proximity, population functional architecture, or both. This question
is important because autism-related differences may extend across coordinated
systems whose regions are functionally related even when they are anatomically
distant. The proposed model represents this organization through anatomical,
diagnosis-blind functional, and interaction kernels within a common Bayesian
hierarchy.

To evaluate the contribution of the covariance construction, we compare three
models fitted to the same participant-level responses and covariates. The
independent model uses
\[
\operatorname{Cov}(\bm\beta_{\mathrm{ASD}})=\tau_I^2I_q
\]
and estimates no structured dependence across connections. The anatomy-only
model uses
\[
\operatorname{Cov}(\bm\beta_{\mathrm{ASD}})=\sigma_A^2K_A,
\]
while the proposed Edge-Space model uses
\begin{equation}
\operatorname{Cov}(\bm\beta_{\mathrm{ASD}})
=
\sigma_A^2K_A+
\sigma_G^2K_G+
\sigma_{AG}^2K_{AG}.
\label{eq:application_covariance}
\end{equation}
The comparison separates the gain from anatomical borrowing from the additional
information contributed by functional organization and its interaction with
anatomy.

For each edge $e$, posterior inference is summarized by its posterior mean,
credible interval, and directional probability
\begin{equation}
\pi_e
=
\max
\left[
P\{\beta_{\mathrm{ASD}}(e)>0\mid Y\},
P\{\beta_{\mathrm{ASD}}(e)<0\mid Y\}
\right].
\label{eq:application_sign_probability}
\end{equation}
Connections with $\pi_e\geq0.975$ are described as posterior supported. This
threshold summarizes directional certainty, while the complete posterior
distribution remains available for every connection.

\subsection{Whole-connectome ASD effect and posterior precision}
\label{subsec:application_effect}

\begin{figure}[p]
\centering
\includegraphics[width=0.92\textwidth]{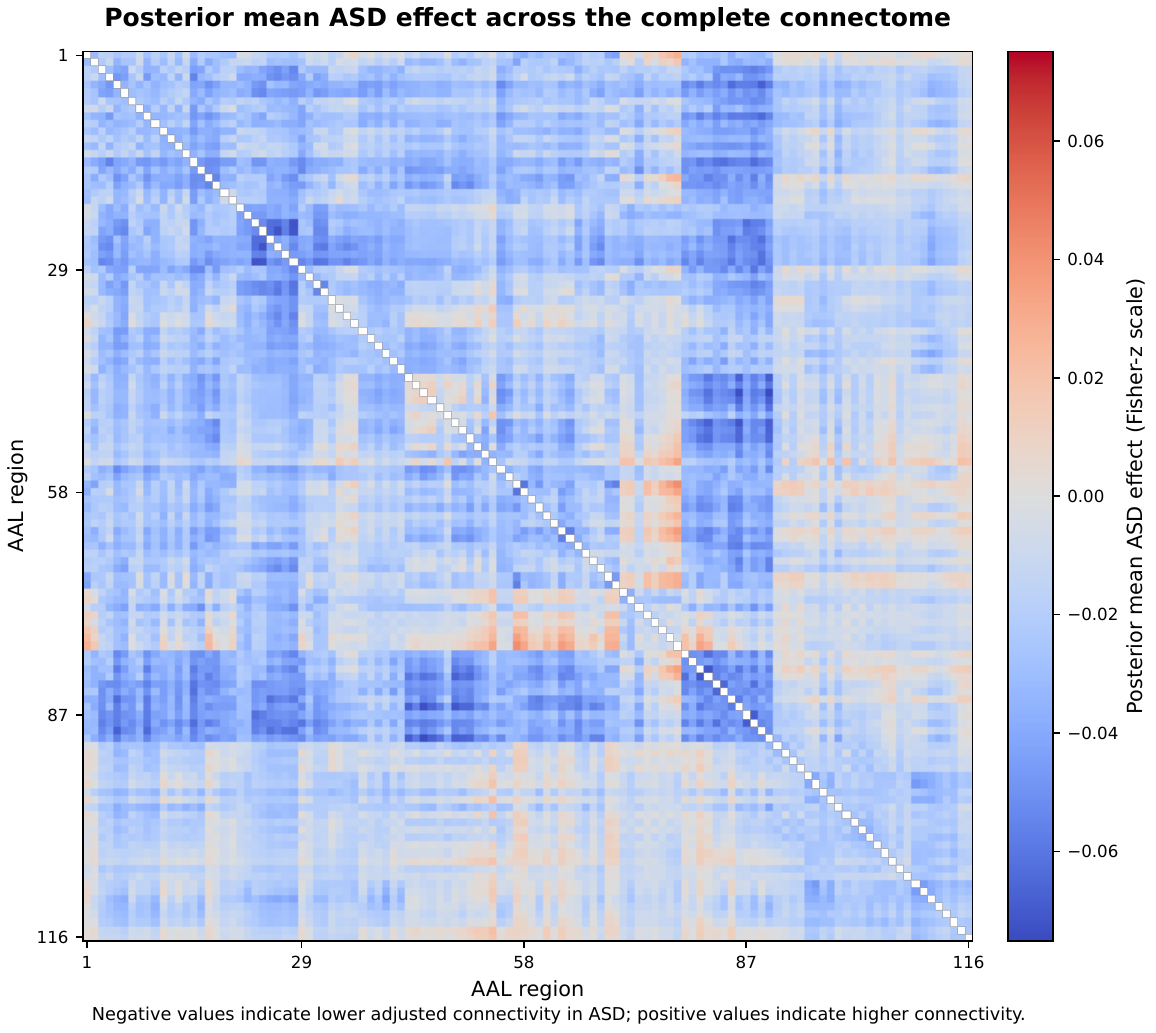}
\caption{
Posterior mean ASD effect across the complete functional connectome under the proposed Edge-Space model. Each off-diagonal entry represents one undirected AAL-region pair. Negative values indicate lower expected connectivity in ASD after adjustment for age, sex, head motion, and acquisition site, whereas positive values indicate higher expected connectivity. The broad negative regions and more localized positive regions show that the ASD-associated pattern is distributed and heterogeneous rather than uniformly increased or decreased.
}
\label{fig:asd_effect_heatmap}
\par\noindent\textbf{Alt text:} Heat map of posterior mean ASD effects across all AAL-region pairs, with predominantly negative values and localized positive regions.
\end{figure}

Figure~\ref{fig:asd_effect_heatmap} displays the posterior mean ASD effect for
all 6,670 connections. Estimates range from $-0.0751$ to $0.0411$ on the
Fisher-$z$ scale. Negative effects are more extensive and generally larger in
magnitude, although localized positive regions remain visible within the broader
pattern.

The results do not support a uniform characterization of autism as either
globally hypoconnected or hyperconnected. Instead, they indicate heterogeneous
reorganization across distributed neural systems, with widespread reductions
and more localized increases occurring in different portions of the functional
architecture. The extended regions of similarly signed effects also suggest
that the observed differences are coordinated rather than confined to isolated
region pairs.

The main inferential contribution is not simply the number of posterior-supported
connections. The Edge-Space model estimates the complete ASD effect surface
while borrowing information across connections that are related through
biologically interpretable geometries. Web Figure 3 in the Supplementary Materials shows the resulting change in posterior precision. The mean posterior standard
deviation decreases from 0.01336 under independent estimation to 0.00351 under
the anatomical model and 0.00337 under the full Edge-Space model. The
corresponding mean 95\% credible-interval widths are 0.05231, 0.01373, and
0.01319.

Under the full model, 5,688 connections meet the posterior sign-probability
criterion, including 5,398 negative and 290 positive effects. These counts
reflect greater directional certainty across the fitted surface rather than a
change in the inferential threshold. The predominance of negative effects
indicates that lower adjusted connectivity is the dominant pattern in these
data, while the positive effects remain localized and structured.

Most of the reduction in average posterior uncertainty arises from
anatomical borrowing. The full Edge-Space model provides a smaller additional
reduction, from 0.00351 to 0.00337, after functional and interaction structure
are added. The similarity between the anatomical and full-model uncertainty distributions in Web Figure 3 should not be interpreted as showing that the full model is simply a more precise version of the anatomical model. Average posterior standard deviation summarizes the overall level of
uncertainty but does not describe how the estimated effects are organized
across the connectome.

The main additional contribution of the full model is structural and
scientific. It represents dependence among connections whose endpoints occupy
similar positions in population functional organization, including connections
between anatomically distant regions. This information is not available from
the anatomy-only model. Section~\ref{subsec:application_structure} shows that
the functional component has the largest posterior structural scale under the
fitted parameterization and contributes broad organization to the estimated
ASD effect surface.

Posterior precision is not uniform across the connectome. Connections supported by stronger anatomical or functional relationships are estimated more precisely, while those with less support retain greater uncertainty. The model therefore adapts borrowing to the organization of edge space rather than applying uniform smoothing; Web Figure 4 displays the connection-specific uncertainty surface.

\subsection{Biological organization of the ASD effect}
\label{subsec:application_structure}

The Edge-Space model adds a structural interpretation to the connection-specific ASD effects. By decomposing the covariance into anatomical, functional, and joint anatomical--functional components, it describes how the estimated effect surface is organized across the connectome.

\begin{figure}[p]
\centering
\includegraphics[width=\textwidth]{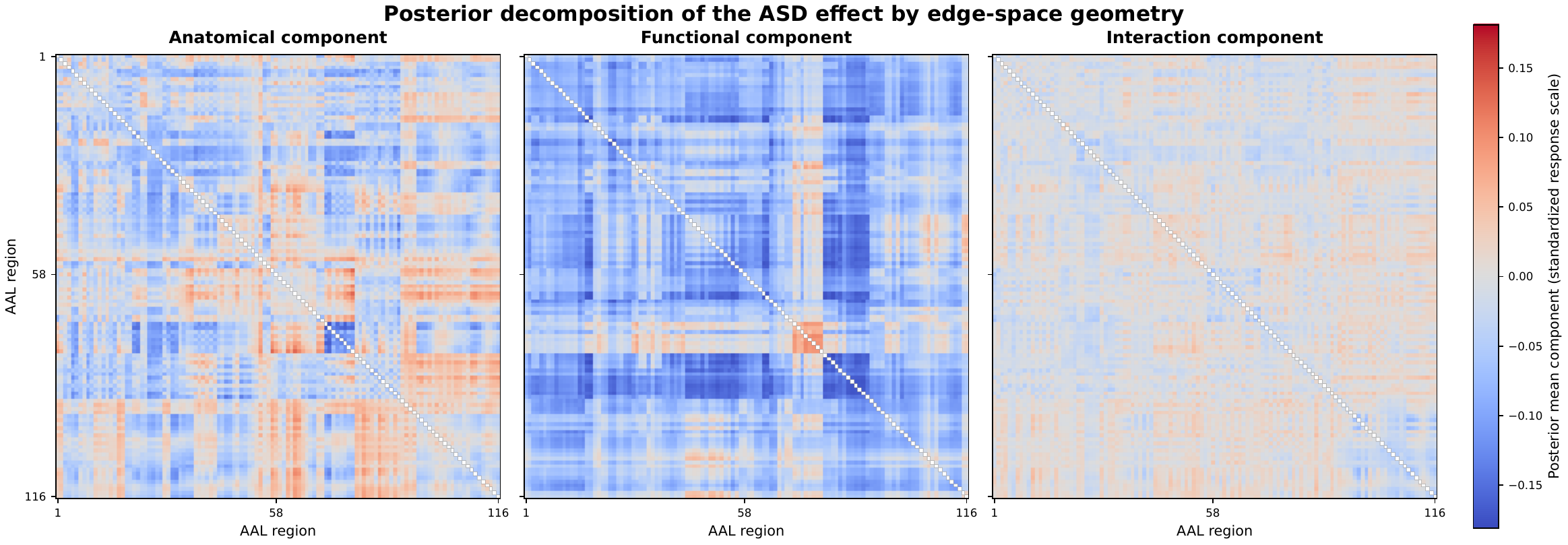}
\caption{
Posterior decomposition of the ASD effect into anatomical, functional, and anatomical--functional interaction components. Values are shown on the standardized response scale used in the fitted hierarchy. The functional component exhibits the strongest and broadest organization, the anatomical component contributes complementary structure, and the interaction captures more localized dependence supported jointly by both geometries.
}
\label{fig:component_decomposition}
\par\noindent\textbf{Alt text:} Three heat maps decomposing the posterior ASD effect into anatomical, functional, and anatomical-functional interaction components.
\end{figure}

Figure~\ref{fig:component_decomposition} displays the posterior mean of each component on the standardized response scale. The functional component is the largest and most coherent, with broad negative regions and several localized positive blocks. The anatomical component contributes additional signed structure, while the interaction component is smaller and more localized.

These results suggest that ASD-associated connectivity differences follow population functional organization beyond what can be explained by anatomical proximity alone. Connections may exhibit related effects when their endpoints occupy similar positions in large-scale functional systems, even when the corresponding regions are anatomically distant. Physical organization remains relevant, and the interaction component captures additional structure supported by both representations.

Table~\ref{tab:structural_scales} quantifies these contributions. Because each retained kernel was normalized after spectral truncation to have average diagonal equal to one, the scales are comparable under the fitted parameterization. The posterior mean is 0.102 for the functional component, 0.042 for the anatomical component, and 0.023 for their interaction.

The larger functional scale is an important scientific result of the model. Independent edgewise analysis can estimate effects at individual connections, but it cannot reveal how those effects are related across the connectome. An anatomy-only model captures local spatial organization but cannot represent dependence associated with distributed functional systems. The full Edge-Space model retains connection-specific inference while learning how anatomical and functional structure support the complete ASD effect surface.

The components should be interpreted jointly and conditionally on the selected representations rather than as uniquely separable biological mechanisms. Their relative magnitudes describe the organization of the estimated effect surface under the fitted model.

\begin{figure}[p]
\centering
\includegraphics[width=1.02\textwidth]{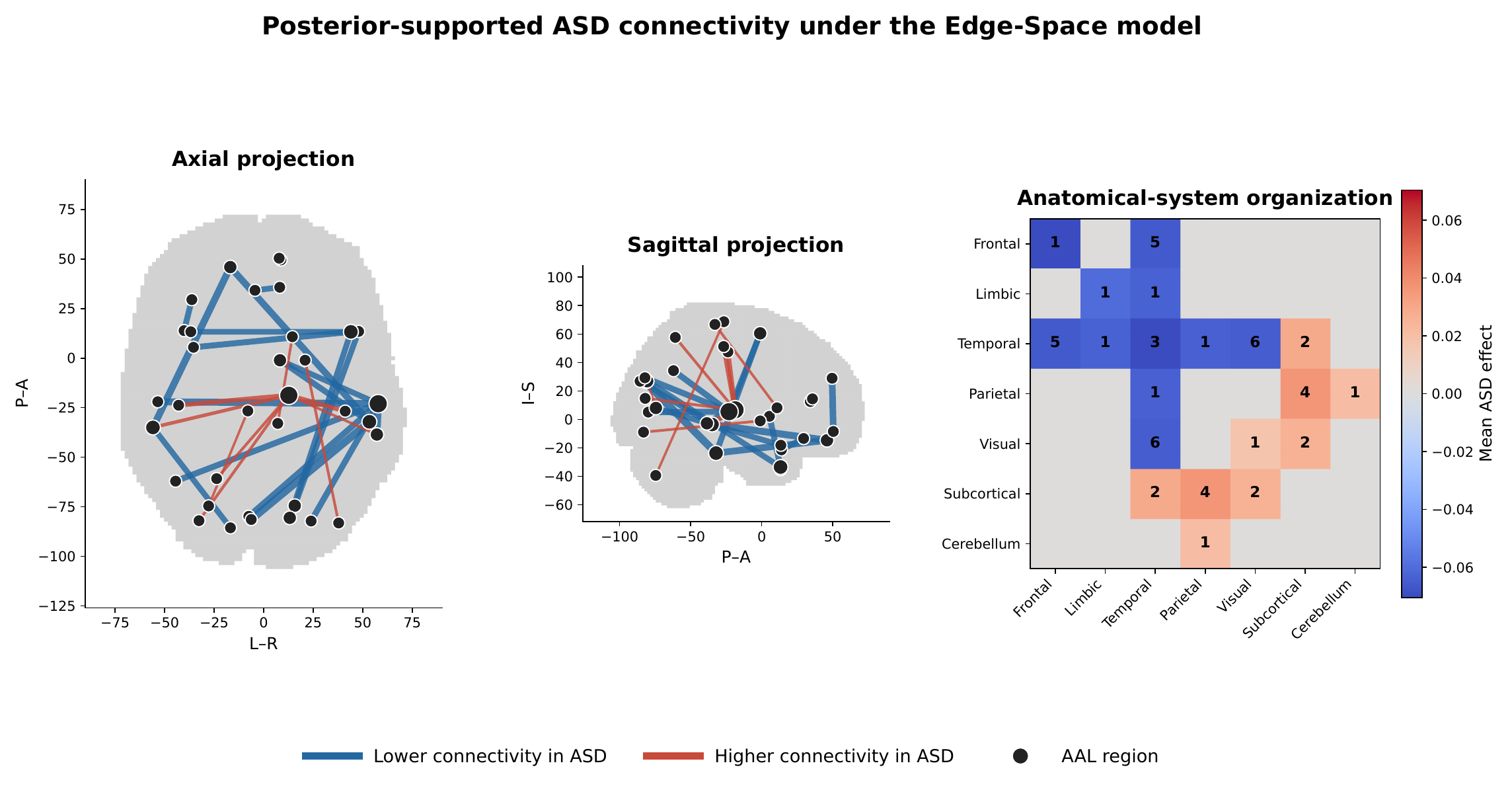}
\caption{
Strongest posterior-supported ASD connectivity differences under the proposed
Edge-Space model. The axial and sagittal projections display the 30 connections
with the largest absolute posterior mean effects among connections with
posterior sign probability at least $0.975$. Blue edges indicate lower adjusted
connectivity in ASD and red edges indicate higher adjusted connectivity; line
width is proportional to the absolute posterior mean effect. Node size reflects
the number of displayed connections incident on each AAL region. The right
panel summarizes the anatomical-system organization of the displayed network.
Cell values give the number of connections between systems, while color
represents their mean posterior ASD effect.
}
\label{fig:posterior_supported_network}
\par\noindent\textbf{Alt text:} Network display of posterior-supported ASD
connections, showing widespread negative and fewer localized positive effects
among AAL regions.
\end{figure}

Figure~\ref{fig:posterior_supported_network} links these whole-connectome results to recognizable neuroanatomical systems. The strongest posterior-supported differences extend across frontal, temporal, limbic, and parietal structures rather than concentrating in one anatomical region. Negative effects predominate, with fewer localized positive effects within the broader pattern of reduced connectivity.

The application shows that the proposed method contributes more than improved precision. It identifies where ASD-associated connectivity differences occur and reveals how those differences are organized across anatomical and functional brain architecture. This structural interpretation supports a view of autism as heterogeneous reorganization across distributed neural systems rather than a change confined to isolated connections or local anatomical neighborhoods.

\section{Discussion}
\label{sec:discussion}

This work introduces a Bayesian framework for whole-connectome inference on the
domain of unordered functional connections. Its methodological contribution is
a positive-semidefinite covariance construction that transfers anatomical and
functional similarities from regions to edges while preserving endpoint
identity, invariance to endpoint ordering, and direct posterior inference at
every connection. Theorem~\ref{thm:edge_space_psd} provides the mathematical
foundation for this construction and extends covariance modeling to settings in
which the inferential units are relationships rather than individual entities.

The framework occupies a distinct inferential position between independent
edgewise analysis and lower-dimensional network methods. It retains adjusted,
connection-specific effects and their uncertainty while learning how those
effects are related across the connectome. The Bayesian hierarchy estimates the
anatomical, functional, and interaction scales jointly with the regression and
site-adjustment components, conditional on the selected kernels.

The simulations in Web Appendix C show that Edge-Space borrowing improves recovery of the complete effect surface, particularly when individual connection signals are weak. The gains in RMSE and correlation support sharing information across connections through biologically informed covariance structure.

In ABIDE, the estimated ASD effect includes widespread reductions in adjusted
connectivity together with more localized increases. This pattern is consistent
with heterogeneous reorganization across distributed neural systems rather than
uniform hyper- or hypoconnectivity. The strongest posterior-supported
differences extend across frontal, temporal, limbic, and parietal structures,
rather than concentrating in one anatomical neighborhood.

The covariance decomposition adds a biologically meaningful layer of
interpretation. Anatomical borrowing accounts for most of the reduction in
posterior uncertainty, while diagnosis-blind functional organization contributes
additional structure beyond physical proximity. Under the fitted kernel
parameterization, the functional component has the largest structural scale,
suggesting that connections can exhibit related ASD effects when their endpoints
occupy comparable positions in large-scale functional systems, even when they
are anatomically distant. These components describe the organization of the
estimated effect surface and are interpreted jointly and conditionally on the
selected representations.

The analysis conditions on fixed anatomical and functional kernels and a
population-level functional geometry. This choice makes posterior inference
feasible for 6,670 connections while retaining the main scientific targets.
Likewise, the common residual variance provides a practical balance between
model flexibility and computational stability. Comparisons among structural
scales remain conditional on the normalized fitted parameterization and should
not be interpreted as invariant measures of biological importance.

The Edge-Space framework identifies where adjusted ASD-associated connectivity
differences occur and reveals how those differences are organized across the
connectome. In ABIDE, it provides a more precise and biologically informative
account of distributed autism-associated connectivity. The broader scientific
contribution is a general framework for inference when the effects of interest
are attached to relationships rather than individual units. This structure
arises in many settings, including gene interactions, social and communication
networks, ecological associations, and linked health outcomes. By extending
covariance modeling to these relationship-valued domains, the proposed
construction opens a path to interpretable, connection-specific inference across
a wide range of complex network data.

\section*{Data Availability Statement}
The data used in this paper to support the findings are publicly available through the Autism Brain Imaging Data Exchange I and the Preprocessed Connectomes Project at \url{https://preprocessed-connectomes-project.org/abide/download.html}. The analysis used the C-PAC preprocessing pipeline, the AAL regional time-series derivative, temporal band-pass filtering from 0.01 to 0.10 Hz, and no global-signal regression. The data are de-identified and openly accessible for research use.

\section*{Supplementary Materials}
Web Appendices, Tables, and Figures, together with supporting code, are available with this paper. The Supplementary Materials contain theoretical details, the complete simulation study, computational diagnostics, kernel-range sensitivity results, and additional graphical displays.

\clearpage
\bibliographystyle{plainnat}
\bibliography{references}

\clearpage
\begin{table}[!htbp]
\centering
\caption{
Comparison of posterior inference for the adjusted ASD effect. Posterior-supported
connections have posterior sign probability of at least 0.975. Mean width is
the average width of the connection-specific 95\% credible intervals.
}
\label{tab:application_model_comparison}
\begin{tabular}{lrrrrrr}
\toprule
Model
& Supported
& Positive
& Negative
& Mean SD
& Mean width
& Range of means \\
\midrule
Independent
& 1,363
& 38
& 1,325
& 0.01336
& 0.05231
& $[-0.0794,\,0.0452]$ \\

Anatomical
& 5,672
& 314
& 5,358
& 0.00351
& 0.01373
& $[-0.0822,\,0.0486]$ \\

Edge-Space
& 5,688
& 290
& 5,398
& 0.00337
& 0.01319
& $[-0.0751,\,0.0411]$ \\
\bottomrule
\end{tabular}
\end{table}

\clearpage
\begin{table}[!htbp]
\centering
\caption{
Posterior summaries for the structural scales in the proposed Edge-Space model.
Intervals are 89\% equal-tail posterior intervals.
}
\label{tab:structural_scales}
\begin{tabular}{lccccc}
\toprule
Component & Mean & SD & 89\% interval & Bulk ESS & $\widehat R$ \\
\midrule
Anatomical $\sigma_A$
& 0.042
& 0.0021
& $[0.039,\,0.046]$
& 1,200
& 1.00 \\

Functional $\sigma_G$
& 0.102
& 0.0110
& $[0.085,\,0.120]$
& 570
& 1.00 \\

Interaction $\sigma_{AG}$
& 0.023
& 0.0021
& $[0.020,\,0.026]$
& 1,300
& 1.00 \\
\bottomrule
\end{tabular}
\end{table}

\clearpage

\clearpage

\clearpage

\clearpage

\end{document}